\documentclass{article}

\usepackage{arxiv}

\usepackage[utf8]{inputenc} 
\usepackage[T1]{fontenc}    
\usepackage{hyperref}       
\usepackage{url}            
\usepackage{booktabs}       
\usepackage{amsfonts}       
\usepackage{nicefrac}       
\usepackage{microtype}      
\usepackage{lipsum}		
\usepackage{graphicx}
\usepackage{doi}
\usepackage{xspace}
\usepackage{xcolor}

\title{Sobolev Sampling of Free Energy Landscapes.}

\author{
    Pablo F. Zubieta~Rico\thanks{\itshape
        Pritzker School of Molecular Engineering,
        The University of Chicago,
        Chicago, IL 60615, USA
    } \\
	\texttt{pzubieta@uchicago.edu} \\
	\And
	Juan J. de~Pablo*\thanks{\itshape
        Materials Science Division,
        Argonne National Laboratory,
        Argonne, IL 60439, USA
    } \\
	\texttt{depablo@uchicago.edu} \\
}

\renewcommand{\headeright}{Preprint}
\renewcommand{\undertitle}{Preprint}
\renewcommand{\shorttitle}{Sobolev Sampling of Free Energy Landscapes}

\hypersetup{
    pdftitle={Sobolev Sampling of Free Energy Landscapes},
    pdfsubject={Preprint},
    pdfauthor={Pablo F. Zubieta~Rico, Juan J. de~Pablo},
    pdfkeywords={Enhanced sampling, Second keyword, More},
}

\newcommand{\ABF}{\textls{\scshape abf}\xspace}
\newcommand{\ANN}{\textls{\scshape ann}\xspace}
\newcommand{\CFF}{\textls{\scshape cff}\xspace}
\newcommand{\FUNN}{\textls{\scshape funn}\xspace}
\newcommand{\SIREN}{\textls{\scshape siren}\xspace}
\newcommand{\PySAGES}{\textls{\textsc{p}y\textsc{sages}}\xspace}
\newcommand{\SSAGES}{\textls{\scshape ssages}\xspace}

\begin{document}
\maketitle

\begin{abstract}
A family of fast sampling methods is introduced here for molecular simulations of systems having rugged free energy
landscapes. The methods represent a generalization of a 
strategy consisting of adjusting a model for the free energy as a function
of one- or more collective variables as a simulation proceeds. Such a model is gradually built as a system evolves through phase space from both the frequency of visits
to distinct states and generalized force estimates corresponding to such states.
A common feature of the methods is that the underlying functional models and their
gradients are easily expressed in terms of the same parameters, thereby providing faster and more effective fitting of the model from simulation data than other available
sampling techniques. They also eliminate the need to train simultaneously separate neural networks,
while retaining the advantage of generating smooth and continuous functional
estimates that enable biasing outside the support grid. Implementation of the methods is relatively simple and, more importantly, they are found to provide gains of up to several orders of magnitude in computational efficiency over existing approaches.
\end{abstract}


\section{Introduction}
Simulations of systems whose underlying free energy landscape consists of local minima separated by large barriers generally require the use of advanced sampling techniques. The Umbrella Sampling method is perhaps the most widely used example\,\cite{torrie1977nonphysical}, where
harmonic restraints are used to coerce the system of interest to visit target states. In order to collect biased statistics, multiple simulations are performed,
subject to constraints along one or more collective variables. The results of these calculations are then combined through a
weighted histogram analysis approach\,\cite{kumar1992weighted} to arrive at a more comprehensive picture of a particular region of the free energy landscape.

A different class of adaptive techniques, which we refer to as density-of-states (DOS) sampling methods\,
\cite{yan2002density,singh2012density},
seek to generate a running estimate of the free energy, on the fly, by relying on visits to distinct points in phase space.
That estimate allows an algorithm to ``push'' a system towards high-energy states,
which would otherwise by visited with a low probability, and eventually cross a barrier.
Examples include the Wang-Landau method\,\cite{yan2002density}, or configurational-temperature density-of-states sampling\,\cite{yan2003fast,singh2012density}, or weighted ensemble methods \cite{chopra2006improved}.
Another example of such methods is provided by ``metadynamics''\,\cite{laio2002escaping},
where the free energy landscape is represented as a
sum of Gaussians with weights and widths determined dynamically as a simulation advances.
Most of these methods require a certain level of intervention by
the user; to improve performance and reduce reliance on user expertise, several newer, ``automated'' alternatives
have been proposed\,\cite{piana2007bias,barducci2008well,dickson2011approaching,singh2011flux},
including basis function sampling\,\cite{whitmer2014basis},
Green's function sampling\,\cite{whitmer2015sculpting},
variationally enhanced sampling\,\cite{valsson2014variational}, and
artificial neural network (\ANN) sampling\, \cite{sidky2018learning,sevgen2020combined} -- the latter approach being perhaps the most robust of the methods in this category.

A fundamentally distinct class of sampling methods is based on the
Adaptive Biasing Force (\ABF) algorithm\cite{darve2001calculating,
darve2008adaptive}.
In such methods, the free energy landscape of the system is estimated
from its derivatives with respect to a set of collective variables.
Methods based on \ABF tend to exhibit good performance.
Working with them, however, can be cumbersome as they usually lack a
simple means of computing the free energy.
Inspired by the success of \ANN sampling, an analogous scheme based
on \ABF was recently proposed \FUNN\cite{guo2018adaptive}.
A combined strategy that takes advantage of both density-of-states
and \ABF based methods has also been proposed recently -- the
so-called combined force frequency \CFF neural-network-based
method\cite{sevgen2020combined} offers superior performance by
relying on both frequencies and forces.

In this work, we present a set of alternatives to \CFF that enable construction of a free energy landscape
from force and frequency information, but that are more
effective and computationally efficient. We also propose a family of methods that allow
one to use more information from any given simulation and facilitate learning the
free energy landscape of a system. As shown below, the performance of the proposed methods is better than that of previously available techniques.

\section{Methods}
\label{sec:methods}

We begin with a brief review of the idea behind the \CFF method. As mentioned above, \CFF is based on the
idea of using artificial neural networks to learn the free energy by querying both force
and frequency information. It employs two neural networks, one to learn from frequency
contributions to the free energy estimate, as described in the original
\ANN technique, and one to learn from force contributions, as in the \FUNN method. More specifically, 
at every step of a simulation, a generalized force is computed according to
\begin{equation}
	\frac{dA}{d\xi} = -\bigg\langle\frac{d}{dt}(Wp)\bigg\vert_\xi\bigg\rangle
\end{equation}
where $W$ is an arbitrary vector field that satisfies an orthogonality constraint, as explained in
Darve et al.s original work\,\cite{darve2001calculating,darve2008adaptive},
and $p$ is the matrix of atomic momenta.

Both neural networks are simultaneously trained to minimize an objective function of the form
\begin{equation}\label{eq:obj}
	\mathcal{C} = \beta\sum_i\Big[(P_i - Q_i)^2 + (F_i - \dot{Q}_i)^2\Big] + \alpha\sum_j w_j^2
\end{equation}
where $P_i$ and $F_i$ are the $i$-th estimates of the free energy and its derivative. $Q$ and
$\dot{Q}$ are the neural network output and its derivative with respect to the
neural network inputs (the collective variable values sampled on its underlying grid). The rightmost
term is a regularization term included to prevent over-fitting.

The $m$-th layer of each neural network can be expressed as
\begin{equation}\label{eq:layer}
	a_m = \phi_m(a_{m-1}) = f{.}\Big(W_m\ a_{m-1} + b_m\Big)
\end{equation}
where $W_m$ and $b_m$ are the layer weights and biases, respectively.
The activation function $f$ chosen for the implementation of \CFF
(as was also done in \ANN and \FUNN) is the $\tanh$ function,
which is applied to all layers except for the output layer,
for which a linear activation function is used.
The dot after $f$ represents an element-wise application of the function.
The network that trains only on the frequencies of visits to states
provides the biasing during the initial stages of the simulation,
as there tends to be a significant mismatch between force and frequency based estimates
in the early states of a simulation.
Training both networks incurs the costs associated with learning the weights of two separate neural networks.
In addition,
\ANN and \CFF need tuning by the user 
to determine the appropriate network size and sweep frequency for a given system.
As shown below, these issues can be circumvented
by learning a single set of parameters for a simpler model
with less user intervention.

\subsection{\texorpdfstring{\textls{S{\scshape iren}}s}{SIRENs}-based sampling}

If the activation function in (\ref{eq:layer}) is changed from $\tanh$ to $\sin$, one obtains
the so-called sinusoidal representation networks, or {\SIREN}s, which exhibit a series of attractive
mathematical properties. For instance, the derivative of a \SIREN is itself a \SIREN.
Moreover, the \SIREN that results from differentiating another \SIREN with respect to the network
inputs can be evaluated in terms of the same weights and with a shift of ${\pi}/{2}$ on the biases.
For a general overview on the description of {\SIREN}s we refer readers to the
work of Sitzmann, et al.\cite{sitzmann2019siren}.
Specifically, if $\mathcal{A}$ is some $\mathbb{R}^n \mapsto \mathbb{R}$ function,
we represent it as a \SIREN according to:
\begin{equation}
	\mathcal{A}(\xi) = W_n\,\big(\phi_{n-1} \circ \phi_{n-2} \ldots \circ \phi_0\big)(\xi) + b_n
\end{equation}
with layer $m$ defined as before:
\begin{equation}\label{eq:sinlayer}
	\phi_m(x) = \sin{.}\Big(W_m\,x + b_i\Big) \;.
\end{equation}

One can therefore express the gradient of the function $\mathcal{A}$ with respect to the input $\xi$ as
\begin{equation}
	\nabla_\xi\mathcal{A} =
	    \Big(W_0^T\,\phi'(a_0)\Big)\cdot\ldots\cdot
	    \Big(W_{n-1}^T\,\phi'(a_{n-1})\Big)\cdot W_n^T
\end{equation}
where $a_{m-1}$ is the input to the $m$-th layer of the \SIREN, and each layer $\phi'_m$ is the same as $\phi_m$ but with a phase shift
\begin{equation}
	\phi'_m(a) = \sin{.}\Big(W_m\,a + (b_m {\,.+\,} \frac{\pi}{2})\Big) \;.
\end{equation}
This permits the evaluation of both a \SIREN and its gradient in a single pass, which is a particularly convenient feature.

These neural architectures can be used to build a wide range of schemes for learning the free energy of a molecular system. The following examples are representative.
First, if one sets the same objective function (\ref{eq:obj}) as before,
and aims to train the network from both the generalized forces and frequencies, the resulting scheme is similar to \CFF, but now the learning proceeds as follows:
\begin{enumerate}
  \item the unbiased histogram is computed in the same way as for \ANN,
  \item the network is trained on the generalized forces and frequency of visited states
  only for a number of initial training sweeps, for example, until all bins for the
  collective variable grid have been visited at least once,
  \item the learning is then switched to train only on generalized forces
  (this reduces the impact that the training frequency might have).
\end{enumerate}

A second, equally effective variation of the previous procedure (see below),
would be to only train the network on the gradients of the forces.
This is similar in spirit to the previously reported \FUNN method, but allows one to
compute the free energy directly from the same neural network.

Note that {\SIREN}s lend themselves to more general problems with cost functions of the form
\begin{equation}\label{eq:general_obj}
	\mathcal{C}\Big(\xi, A(\xi), \nabla_\xi A(\xi), \nabla_\xi^2A(\xi), \ldots \Big) = 0.
\end{equation}
In principle, one can consider training a model not only using forces and frequencies but,
for certain collective variables, with the gradients of the forces, which can be expressed in terms
of the local configurational temperature, given by
\begin{equation}
	\big\langle\sum_i\nabla\cdot F_i\ \big\rangle =
	    -\frac{1}{k_BT_\mathrm{conf}}\big\langle\sum_i F_i^2\ \big\rangle
\end{equation}

\subsection{Spectral Sampling}
\label{eq:spectral}

The algorithms outlined in the previous section, specifically Equation (\ref{eq:general_obj}), without regard for the functional form of $\mathcal{A}$, are suggestive of a more general family of sampling methods. Indeed, any
function in a Sobolev space for which its values and its derivatives can be conveniently expressed
in terms of the same set of parameters could in principle replace the use of {\SIREN}s.

For low dimensional problems (one or two collective variables),
one might adopt a generalized approach based on Basis Functions Sampling\,\cite{whitmer2014basis},
and model both forces and the free energy as expansions (or tensor product expansions)
in an appropriate set of bases.

To illustrate this idea, consider approximating the free energy as a function of a single
periodic collective variable using a truncated Fourier series
\begin{equation}\label{eq:fourier}
	A(\xi) \approx
	    c_0 + \mathrm{Re}\sum_{k=1}^n c_k z^k =
	    \sum_{k=1}^n a_k \mathrm{Re}z^k - b_k \mathrm{Im}z^k
\end{equation}
where $z = e^{i k \xi}$. It follows that the forces can be expressed as
\begin{equation}\label{eq:dfourier}
	-F(\xi) = \frac{\partial A(\xi)}{\partial\xi} \approx
	    \mathrm{Re}\sum_{k=1}^n i k c_k z^k =
	    -\sum_{k=1}^n k\,a_k\,\mathrm{Im}z^k + k\,b_k\,\mathrm{Re}z^k \;.
\end{equation}

If we collect partial statistics of the forces, as is done in \ABF, we can fit the coefficients
of the expansion by means of an \emph{alternant matrix} for the $m$ support points of the
helper histogram
\begin{equation}
	-\mathrm{Im}\begin{pmatrix}
        z_1    & 2z_1^2 & \ldots & nz_1^n \\
        z_2    & 2z_2^2 & \ldots & nz_2^n \\
        \vdots &        &        & \vdots \\
        z_m    & 2z_m^2 & \ldots & nz_m^n
    \end{pmatrix}
    \begin{pmatrix}
        a_1    \\
        \vdots \\
        a_n
    \end{pmatrix}
    -
    \mathrm{Re}\begin{pmatrix}
        z_1    & 2z_1^2 & \ldots & nz_1^n \\
        z_2    & 2z_2^2 & \ldots & nz_2^n \\
        \vdots &        &        & \vdots \\
        z_m    & 2z_m^2 & \ldots & nz_m^n
    \end{pmatrix}
    \begin{pmatrix}
        b_1    \\
        \vdots \\
        b_n
    \end{pmatrix}
    \approx
    \begin{pmatrix}
        F_1    \\
        \vdots \\
        F_m
    \end{pmatrix},
\end{equation}
and, for example, solve the latter expression as a least squares problem.

Since the coefficients are the same for the expansion of the energy and the approximation
of the force (except for the $c_0$), one also obtains an expression for the free energy up to an additive
constant. What we have just described is in reality an alternative to the \FUNN
sampling method that uses basis expansions instead of a neural network,
and the idea can be generalized to parallel the \CFF approach as well.
Consider the following Vandermonde and diagonal matrices
\begin{equation}
	M = \begin{pmatrix}
        z_1    & z_1^2 & \ldots & z_1^n  \\
        z_2    & z_2^2 & \ldots & z_2^n  \\
        \vdots &       &        & \vdots \\
        z_m    & z_m^2 & \ldots & z_m^n
    \end{pmatrix},\qquad
    \mathcal{D} = \begin{pmatrix}
        1      & 0 & \ldots & 0 \\
        0      & 2 &   & 0 \\
        \vdots &   & \ddots & \vdots \\
        0      & 0 & \ldots & n
    \end{pmatrix}.
\end{equation}
By also collecting the frequencies, as in \CFF, and estimating the free energy $\{A_1, \ldots, A_n\}$
from the corresponding re-weighted histogram, as in \ANN, we can now solve
\begin{equation}
    \begingroup
    \renewcommand*{\arraystretch}{1.5}
    \begin{pmatrix}
        \;1  & \mathrm{Re}(M) \\
        \;0  & -\mathrm{Im}(\mathcal{D}M)
    \end{pmatrix}
    \endgroup
    \begin{pmatrix}
        c_0    \\
        a_1    \\
        \vdots \\
        a_n
    \end{pmatrix}
    -
    \begingroup
    \renewcommand*{\arraystretch}{1.5}
    \begin{pmatrix}
        \;\mathrm{Im}(M)\; \\
        \;\mathrm{Re}(\mathcal{D}M)\;
    \end{pmatrix}
    \endgroup
    \begin{pmatrix}
        b_1    \\
        \vdots \\
        b_n
    \end{pmatrix}
    \approx
    \begin{pmatrix}
        A_1    \\
        \vdots \\
        A_m    \\
        F_1    \\
        \vdots \\
        F_m .
    \end{pmatrix}
\end{equation}

Note that the spectral approach just described will not scale as easily to a large number
of CVs as the neural network approaches but, given the abundance of systems where only one or two CVs
are important, this method should yield faster sampling convergence, because
the costs of fitting the spectral approximation should in general be lower than the computational
work required to train neural networks for low-dimensional collective variable spaces.

The set of methods described in this section can also be viewed as applications of different techniques
from \emph{approximation theory}\,\cite{trefethen2019approximation}.
For instance, if the chosen bases are ultra-spherical polynomials,
the scheme would be closely related to the well-conditioned spectral method by Olver and Townsend\,\cite{olver2013fast} 
for approximating functions and solving ordinary differential equations\,\cite{townsend2015automatic}.
A useful result is that a least-squares polynomial approximant of degree $n$
that satisfies $n \le \frac{1}{2}\sqrt{m}$\,\cite{demanet2019stable},
$m$ being a number of equally spaced sampling points,
provides weak guarantees on the stability of extrapolations near the sampling domain boundaries.
We take this as a guiding principle to automatically choose the polynomial order from the number of points of a user-defined grid for the collective variable.


\section{Results}
\label{sec:results}

In all simulations we followed the same overall strategy to evaluate our approximate
model for the free energy as in previous neural network-based sampling methods. We train or fit the
model every $N$ time steps of a simulation, and use the smooth continuous function represented by the
selected model to provide and estimate of the forces to bias the simulation.
Unless stated otherwise, all methods were implemented and ran with the \PySAGES
(Python Software Suite for Advanced General Ensemble Simulations) software package,
which is a Python implementation of the \SSAGES library\,\cite{sidky2018ssages}
designed for advanced sampling simulations on \textls{GPU}s.

\subsection{Alanine Dipeptide in Water}

\begin{figure}[htbp]
\centering
\includegraphics[width=0.45\textwidth]{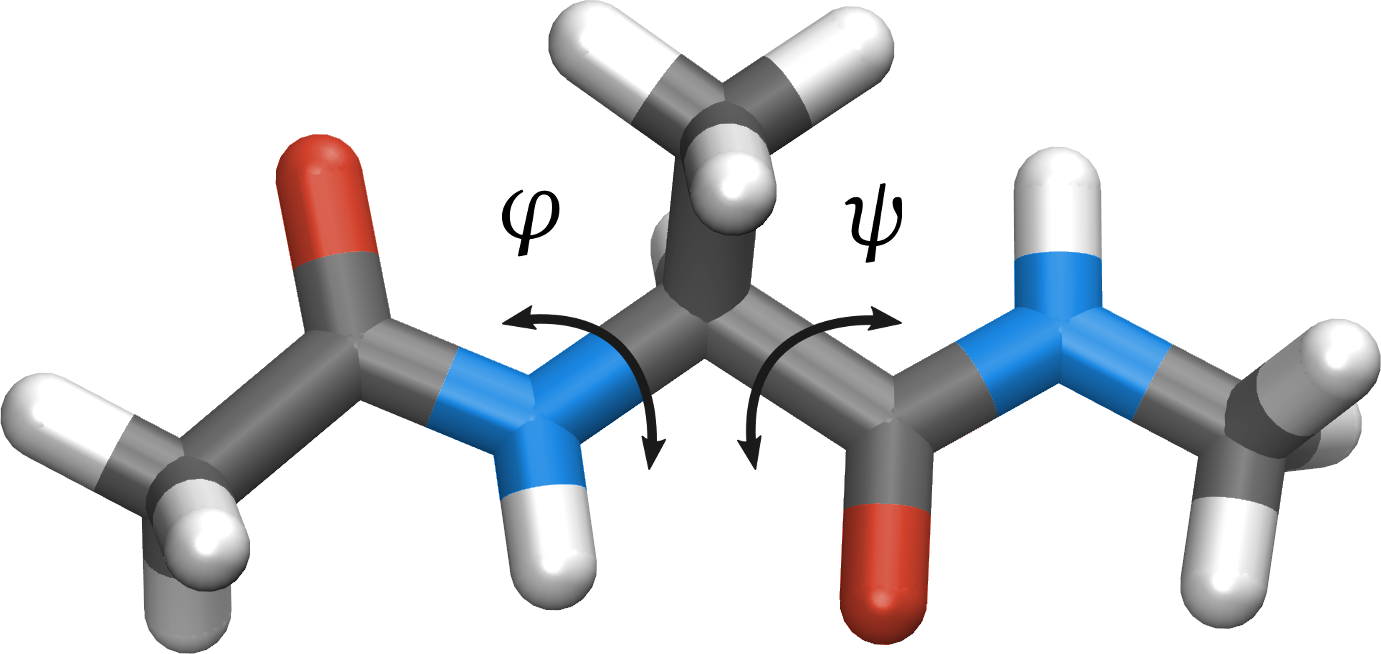}
\caption{Dihedral angles used as canonical CVs to describe alanine dipeptides}
\label{fig:ala}
\end{figure}

To examine the performance of the algorithms introduced above we start with one of the canonical examples that is used to benchmark enhanced sampling methods, namely the alanine dipeptide (Figure \ref{fig:ala}) in explicit water.
For this system, simulations were performed using OpenMM 7.5\,\cite{eastman2017openmm} as a back-end
for \PySAGES with the Amber99sb force field and 750 \textls{TIP3P} water molecules,
in a box of size 3~nm $\times$ 3~nm $\times$ 3~nm.
Simulations were performed at $T = 298.15~\mathrm{K}$.
The equations of motion were integrated using a Langevin thermostat with a friction
coefficient of $1~\mathrm{ps}^{-1}$ and a 2~fs time step.

In what follows we compare the following sampling methods:
\begin{enumerate}
  \item A \SIREN network with a single inner layer with 14 nodes, trained with forces and frequencies on a collective variable grid of $32\times32$,
  \item a tensor-product Fourier expansion fitted on the forces, and
  \item a tensor-product Fourier expansion fitted on both frequencies and forces.
\end{enumerate}
For both spectral (Fourier) sampling methods, we sample over a $64\times64$ grid.
In addition we use as a reference a 100~ns \ABF sampling run
and a (12--8) \CFF network run with \SSAGES and Gromacs\,\cite{van2005gromacs},
as described in the original \CFF publication.

\begin{figure}[htbp]
\centering
\includegraphics[width=0.75\textwidth]{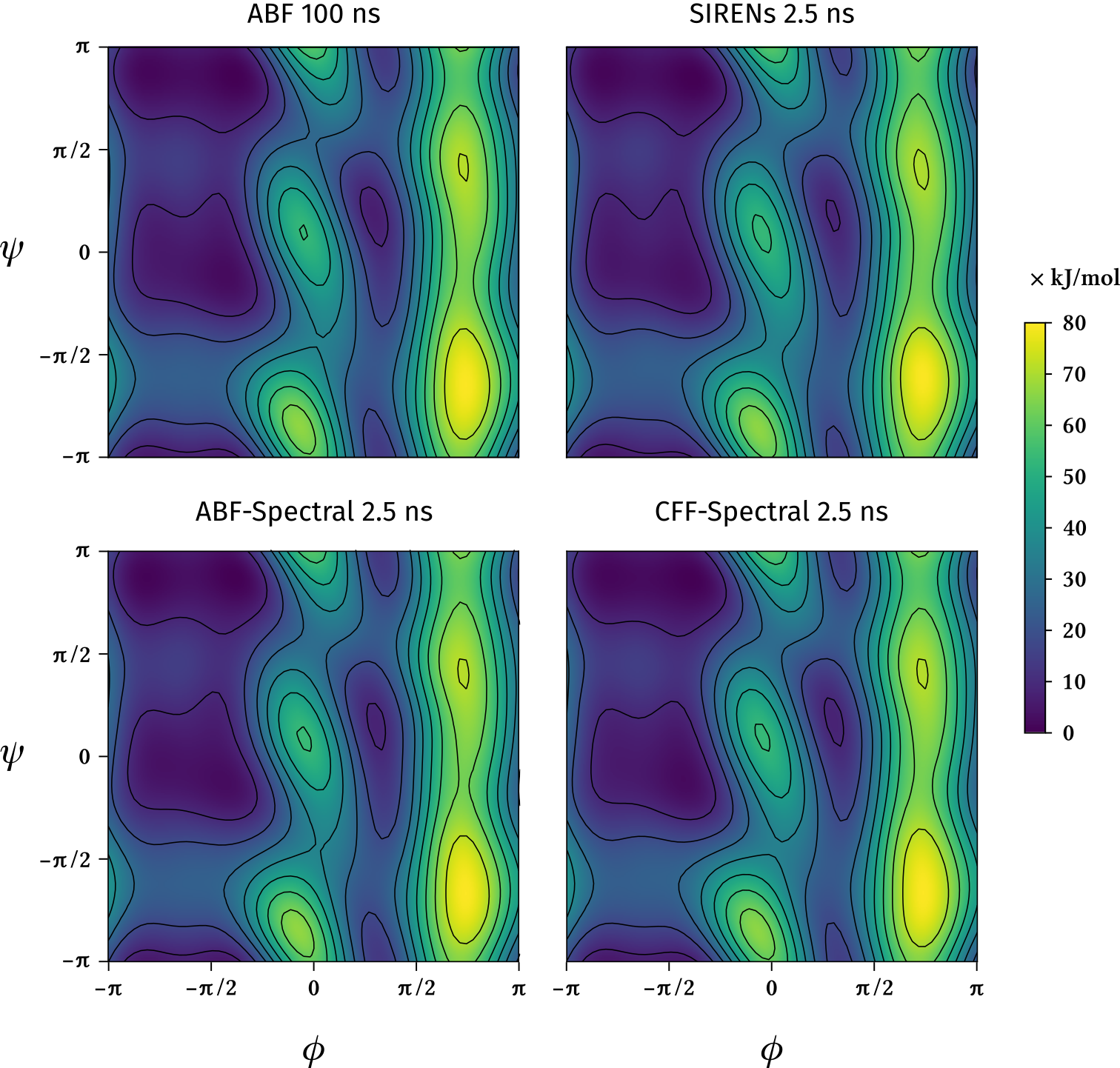}
\caption{
  Comparison of the free energy landscape of alanine dipeptide
  as a function of the dihedral angles from Fig.~\ref{fig:ala} for the
  methods introduced in this work. A 100~ns run of \ABF is plotted for reference.
}
\label{fig:ala-pmf}
\end{figure}

\begin{figure}[htbp]
\centering
\includegraphics[width=0.65\textwidth]{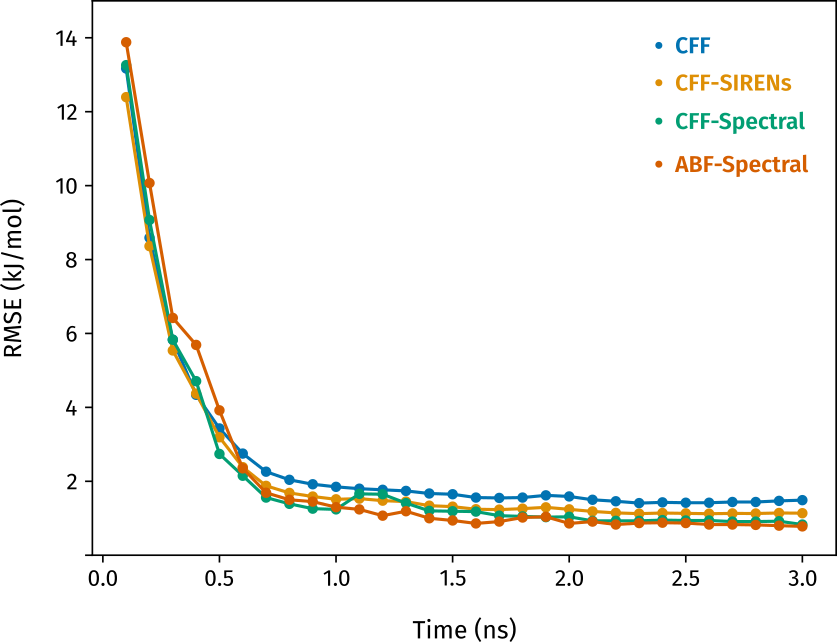}
\caption{Root mean square error of the free energy estimates for both methods}
\label{fig:convergence}
\end{figure}

Analysis of the convergence of the {\SIREN}s-based \CFF and spectral sampling methods reveals the same overall behaviour. This suggests that the general underlying strategy of relying on a single parametric model,
and computing its parameters using frequency and forces estimates,
leads to comparable explorations of the CV space.
At 2.5~ns all methods recover with good accuracy the full energy landscape
(Figure \ref{fig:ala-pmf}).
However, all methods proposed here converge faster than \CFF
(Figure \ref{fig:convergence}).
In particular, we believe that the global nature of the spectral approximations
results in better bias estimates across the full sampling domain,
whereas activated neural network approaches can lead to improvements
closer to that of visited sites during a training sweep and the previous one.

Another important advantage of the \SIREN method presented here over \CFF,
and of all spectral methods over neural based approaches,
is the reduced computational cost.
For instance, training {\SIREN}s was found to be at least 5 times faster
than training $\tanh$-based networks of the same size.
Even more striking is that a mere 14-node single layer perceptron was able to
learn the free energy landscape for this example.

For both spectral methods, fitting every $500$ integration steps leads to
simulations of around 850 \textls{TPS}
on a single Nvidia GeForce \textls{RTX} 2070 Super \textls{GPU},
whereas the {\SIREN} runs with training every $5000$ steps proceeded at nearly 150 \textls{TPS}.
The spectral approach generates a converged free energy landscape
for the alanine dipeptide in water after approximately 15 minutes on the above hardware.

\subsection{One-Dimensional Rugged Free Energy Landscape}

To further illustrate the robustness and convergence of the methods proposed here,
we now consider a one-dimensional, artificial rugged free energy landscape
constructed by adding 100 Gaussians.
This landscape can be viewed as representative of that of a
protein folding funnel or a glassy system.
It is worth noting that this is a more complex surface than the one previously used
to evaluate both the Green-Function Sampling and the \ANN methods, where
the sum of 50 Gaussians was used.

For this example, simulations were carried out with \PySAGES
using HOOMD-blue~2.9.6\,\cite{anderson2020hoomd} as the back-end.
A single 1~amu particle restrained to
one dimension in a simulation box of 10~nm was integrated using Langevin dynamics
with $dt = 0.0005$ and $kT = 1$. The external potential consists of the sum
of 100 Gaussians with heights, widths and centers randomly chosen from the
distributions $\mathcal{U}(-20, 5)$, $\mathcal{U}(1/1000,1/\sqrt{20})$,
and $\mathcal{U}(-4, 4)$, respectively.

\begin{figure}[htbp]
\centering
\includegraphics[width=0.65\textwidth]{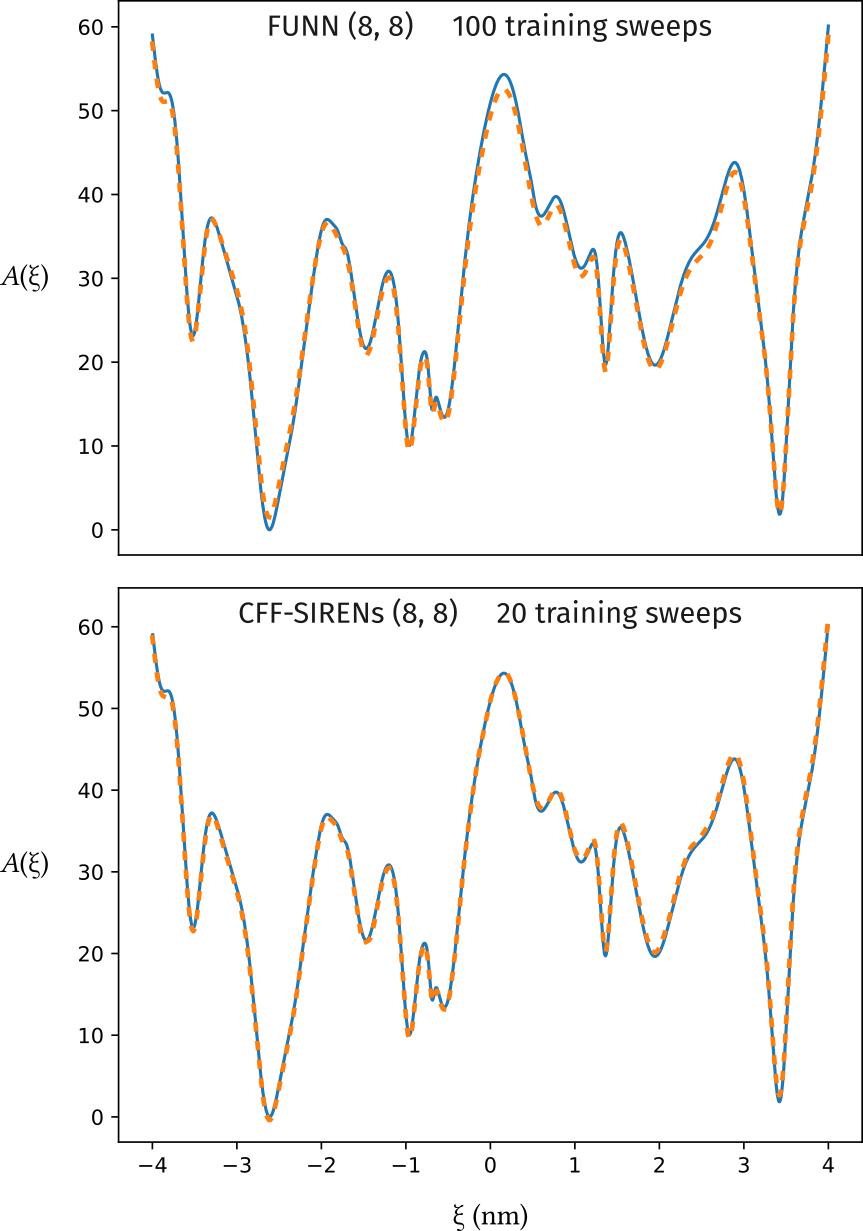}
\caption{Matching the sum of 100 gaussians with {\SIREN}s and \FUNN}
\label{fig:rugged}
\end{figure}

The potential was sampled with a two-layer \SIREN network with two
hidden layers of eight nodes each (8, 8)
at a training frequency of 5000 integration steps.
As a means to assess the performance of the method, we use a \FUNN network with the
same architecture. 
We can observe (Figure \ref{fig:rugged}) that after 20 training sweeps,
the {\SIREN}s sampling scheme is able to basically recover the full potential.
In contrast, after 100 training sweeps,
the \FUNN method has qualitatively learned all of the landscape features,
but it still fails to match the heights of most of the hills by about 3~$kT$.

\subsection{Stretching of Deca-alanine}

As a final example, we compare the spectral method proposed here to other recent methods from the
literature that also try to combine learning from frequencies and forces,
such as the \textls{GaWTM--eABF} hybrid method\,\cite{chen2021overcoming},
which combines Well-tempered Metadynamics
with the extended \ABF method and Gaussian-accelerated molecular dynamics
strategies.
To do this, we consider the stretching of deca-alanine reported in the \textls{GaWTM--eABF}
work. All simulations were again performed with \PySAGES and OpenMM 7.5 as back-end.
All bonded and non-bonded interactions were modeled with the \textls{C\textsc{harmm}22} force field\,\cite{mackerell1998all}.
The temperature was set to 300~K and the system was evolved using a Langevin integrator ($dt = 0.5$~fs)
after a standard energy minimization step.

The collective variable used to bias the simulation was the same as for the \textls{GaWTM--eABF} work,
namely the distance between the first and the last $\alpha$-carbon atoms of the peptide.
For this, the CV domain (4~nm--32~nm) was split into two Chebyshev-distributed sets of points
4~nm--13~nm and 13~nm--32~nm (256 points per region) that we used for binning both forces and frequencies.
The free energy is then modeled as the two piece-wise sets of polynomial expansions.

\begin{figure}[htbp]
\centering
\includegraphics[width=0.65\textwidth]{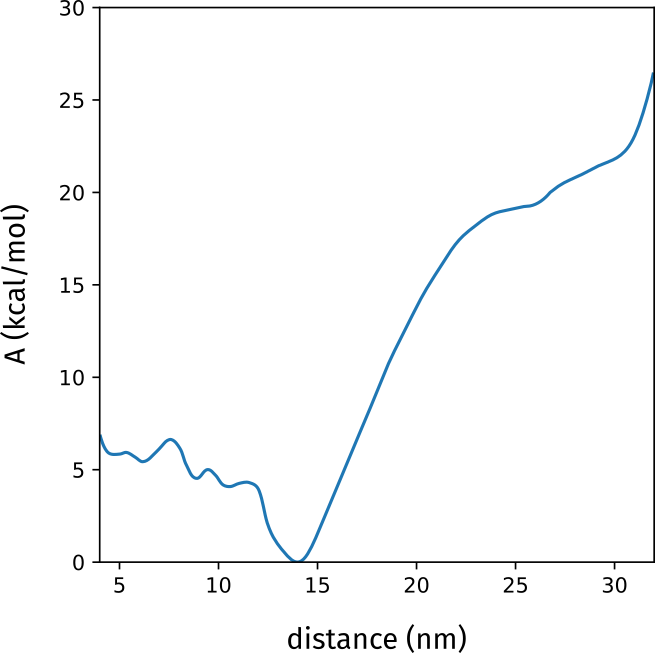}
\caption{Free energy of stretching deca alanine as a function of the distance between terminal $\alpha$ carbons}
\label{fig:deca-ala}
\end{figure}

After a single 50~ns simulation we are able to recover the free-energy profile shown in Figure \ref{fig:deca-ala}.
This free-energy profile portrays all the features of the one obtained by a multiwalker 200~$\mu$s ABF simulation,
and exhibits greater fidelity than the \textls{GaWTM--eABF} strategy, which mixes three different generally robust
sampling methods to bias the system. Furthermore, our proposed approach does not need the a-priori calibration of the method to set any user-facing parameters (other than the size and distribution of the sampling grid).

\section{Conclusions}
\label{sec:conclusions}

A new family of fast and globally convergent enhanced sampling methods
had been proposed. First, building on the idea of learning
free energy surfaces by training neural networks on both frequency of visits
to CV states as well as on generalized forces, we introduced the use of {\SIREN}
neural networks with periodic activation functions that provide a better description of a function and its gradients. The proposed {\SIREN}s are shown to be more stable and efficient
than networks that rely on other activation functions.
Going beyond {\SIREN} sampling, we also proposed two methods that
follow the same fitting strategy, but that are inspired by previously proposed
algorithms based on basis-function sampling. We refer to such methods as Spectral sampling.

The effectiveness of the proposed new algorithms was established through a series of examples. In all cases the performances of {\SIREN}s and Spectral methods
surpass those of all other methods considered here, not only in terms of
convergence, but also from a computational perspective,
even for extremely rugged free-energy landscapes. All the methods proposed have been implemented in the publicly available {\PySAGES} software suite.

\section{Acknowledgements}

This work is supported by the Department of Energy,
Basic Energy Sciences, Materials Science and Engineering Division,
through the Midwest Integrated Center for Computational Materials (\textls{MICCoM}).

\bibliographystyle{unsrt}
\bibliography{ssofel}  






\end{document}